\documentclass[a4paper,10pt]{article}

\usepackage{listings}
\usepackage{amssymb}
\usepackage{amsmath}
\usepackage{amsfonts}
\usepackage{amssymb}
\usepackage{graphicx}
\usepackage{color}%
\usepackage[utf8]{inputenc}

\title{Models for the Propensity Score that Contemplate the Positivity Assumption and their Application to Missing Data and Causality. }
\author{Julieta Molina$^{(1)}$, Mariela Sued$^{(2)}$, Marina Valdora$^{(1)}$, \ \\{\small (1) Universidad de Buenos Aires, (2) Universidad de Buenos Aires and Conicet.}}

\begin{document}

\maketitle

\begin{abstract}

Generalized linear models are often assumed to fit propensity scores, which are used to compute inverse probability weighted (IPW) estimators.
In order to derive the asymptotic properties of IPW estimators, the propensity score is  supposed  to be bounded away from cero. This condition is known in the literature as strict positivity (or positivity assumption) and, in practice, when it  does not hold, IPW estimators are very unstable and have a large variability. 
Although strict positivity is often assumed, it  is not upheld  when some of
the covariates are continuous. In this work,  we attempt to conciliate between the strict positivity condition and the theory of generalized
linear models by incorporating an extra parameter, which results in an explicit lower bound for the propensity scores.

\end{abstract}


\maketitle


\section{Introduction}

In the last 20 years, inverse probability weighted estimators (IPW) have attracted considerable attention in the statistical community.
Among other things, they are used for estimating a population mean $E(Y)$  from an incomplete data
set under the missing at random (MAR) assumption. MAR establishes
that the variable of interest $Y$ and the response indicator $A$ are conditionally independent given an always observed vector $\mathbf X$ of covariates. See References \cite{robinsetal94,robinsetal95,littlean04,kangshafer07}.
In the causal framework IPW estimators are used
for estimating the average effect of a binary treatment on a scalar
outcome under the assumption of no unmeasured confounders  in observational studies. See References \cite{rosenbaum98,hiranoetal03,lunceforddavidian04,crumpetal09,yaoetal10,hill13}.

In the missing data setting, the propensity score is defined as the response probability given the vector of covariates $\mathbf X$ whereas,
 in the causal context, the propensity score is the conditional probability of treatment assignment given a set of measured baseline covariates. IPW estimators are essentially weighted means of observed responses in which the weights are determined as the inverse of the estimated propensity scores. The aim of these weights is to compensate for the missing responses.

One of the conditions required to derive the asymptotic properties of  IPW estimators is the strict positivity condition (also known as the positiviy assumption).
It states, in the missing data setting, that the propensity score is bounded away from zero and,
in the causal context, that the propensity score is bounded away from zero and one (see References \cite{robinsetal95,kangshafer07,lunceforddavidian04,crumpetal09,hill13}).
 Besides theoretical issues, the violation of the strict positivity condition causes the estimates to be very unstable and to have a large variability.
See also References \cite{littlerubin87,colehernan08,leeetal11}.

Most users of IPW procedures assume models for the propensity score that are usually incompatible with
the positivity condition. For instance, this is the case of the generalized linear models (GLM), when some of the covariates are continuous.
Despite this incompatibility, GLM are the most popular models considered in the field. 

In this work we present a slight modification to the GLM originally postulated for the propensity score that is compatible with the strict positivity condition. To explore plausible applications of the new model, we  revise a real example in the causal context. We consider the data collected by  Tager et. al \cite{tageretal79, tageretal83}, who
investigate the effects of cigarette smoking on children's pulmonary function.
They study the effect of parental smoking on the pulmonary function of their children as well as the effect of direct smoking by the children themselves. Besides, in his book,
Rosner \cite{rosner99} performs another analysis of the mentioned data. The author considers the forced expiratory volume (FEV) of a group of children as a response variable as well as their height, age, sex and a binary variable indicating whether or not they smoke. Kahn \cite{kahn05} studies this data using linear regression. In this work we study the effect of smoking in the forced expiratory volume of children by estimating the average treatment effect. This is an example of an observational study in which, under the assumption of no unmeasured confounders, the average treatment effect can be estimated using IPW procedures. The validity of no unmeasured confounders is beyond the scope of this work, and thus, the estimates here presented for the ATE are included only to illustrate the use of the new model for the propensity score. Since the proportion of smokers among the younger children is very small, the corresponding propensity scores take on values near zero and therefore the classical IPW estimators may have a large variance. In situations such as this, we expect that the models we propose in this work will provide more stable estimators than the classical IPW procedures.

The paper is  structured as follows: in
Section 2 we review methods based on propensity scores for estimating a population mean from incomplete data and we present our model for the propensity score in that missing data context. In Section 3 we review methods for estimating average treatment effects that use weighting
by the inverse of the probability of treatment and present our model in the causality context. In Section 4 we describe the computational method we used to estimate the parameters indexing the new model. In Section 5 we study the performance of our model through its impact in IPW estimators by means of a Monte Carlo study. As an illustration, in Section 6, we apply these methods to estimate the effect of smoking in children.

\section{Missing Data }\label{sec:missing}
Assume that we are interested in estimating the mean value of a scalar response $Y$, based on a sample
composed by an always observed vector $\textbf{X} \in  \mathbb{R}^p $ of covariates while the response of interest  is missing by happenstance for some subjects.
We will assume that data are missing at random (MAR) \cite{rubin76}. This means that the  missing mechanism is not related to the response of interest and  it is only
related to $\textbf{X}$, the observed vector of covariates. Let $A$ be a binary variable indicating whether $Y$ is observed or not,
namely, $A=1$ if $Y$ is observed and $A=0$ if $Y$ is missing.
MAR establishes that
\begin{equation}
 \label{MAR}
 P(A=1\mid  \boldsymbol X,Y)=P(A=1\mid \boldsymbol X)=\pi(X).
\end{equation}
$\pi(\boldsymbol X)$ is known in the literature as the propensity score \cite{rosenbaumrubin83}. 
Up to integrability conditions, under the MAR assumption, $\mu=E(Y)$  can be represented in terms of the distribution of the
observed data as $E(Y)=E\left\{AY/\pi(\textbf{X})\right\}$. This representation invites us to estimate $\mu$ by
\begin{equation}
 \label{ipw_missing}
 \hat\mu(\hat \pi) =\mathbb P_n\left\{\frac{AY}{\hat \pi(\textbf{X})}\right\},
\end{equation}
where $\hat\pi(\textbf{X})$ is a consistent estimator of $\pi(\textbf{X})$, and $\mathbb P_n $ is the empirical mean operator, i.e.
$\mathbb P_n V=n^{-1}\sum_{i=1}^n V_i$ \cite{horvitzthompson52}. 
Observed responses corresponding to  low values of the estimated propensity score are highly weighted since they should compensate for the high missing rate associated to such a level of covariates.  For more details see \cite{robinsrotnitzky92,robinsetal94}. 
Different ways of estimating $\pi(\textbf{X})$ will lead to different estimators for $\mu$; parametric and nonparametric estimators of the propensity score have been considered. 
Robins, Rotnitzky and Zhao \cite{robinsetal94} presented a class of  estimators for $\mu$, which containes $ \hat\mu(\hat \pi)$. These estimators are asymptotically normal when $\pi$ follows a parametric model.
 Little and An \cite{littlean04} proposed to estimate the propensity by fitting a
spline to the logistic regression of the missing-data indicator $A$ on $\boldsymbol X$.
Kang and Shafer \cite{kangshafer07} presented $\hat\mu(  \hat \pi)$,  where $\pi$ follows a linear logistic regression model. More generally, parametric models are often assumed for the propensity score and, in particular,  generalized linear models (GLM) play a prominent role  among them.  These models postulate that
\begin{equation}
  \label{para}
  \pi(\textbf{X})=\phi(\boldsymbol{\beta}_0^T \textbf{X}),
 \end{equation}
where  $\phi$ is a strictly increasing cumulative distribution function and $\boldsymbol{\beta}_0 \in  \mathbb{R}^p$. Taking $\phi(u) = 1/\{1 +exp(-u)\} $
results in the linear logistic  model, which is one of the most popular choices in the literature.

Most of the asymptotic results established for
 $\hat \mu(\hat{\pi})$, defined in (\ref{ipw_missing}),
 require the strict positivity condition, which establishes that $P\left\{\pi(\textbf{X})\geq \varepsilon\right\}=1$, for some $\varepsilon >0$ (see References \cite{robinsetal95,kangshafer07}). 
  In practice, when this condition does not hold, values of $\hat \pi (\boldsymbol X_i)$ close to zero may arise causing $\hat \mu(\hat \pi)$ to be a very unstable estimator, having  a large variability ( see also Reference \cite{littlerubin87}). 

Unfortunately,  generalized linear models
prevent the validity of the strict positivity condition, except when $\boldsymbol{\beta}_0^T \textbf{X}$ is bounded from below.
Thus, strict positivity  is typically violated as far as continuous  covariates are included in the vector $\boldsymbol X$.

In this work, we attempt to conciliate between the strict positivity condition and the most popular parametric models used for the propensity score.
To do so, we slightly perturb the original parametric model postulated by the practitioner by incorporating an explicit lower bound for the propensity score. Namely, model (\ref{para})  is replaced by
$$\pi(\textbf{X})=(1-\varepsilon_0)\phi(\boldsymbol{\beta}_0^T \textbf{X})+\varepsilon_0,$$ where $\varepsilon_0 \in (0,1)$  and $\boldsymbol{\beta}_0 \in  \mathbb{R}^p$. This model contemplates the validity of the strict positivity condition and thus we  call it \textit{strictly positive propensity score} (SPPS) model for the missing mechanism. 
Let $\boldsymbol\theta=(\varepsilon, \boldsymbol{\beta}^T)^T$, with $\varepsilon \in (0,1)$  and $\boldsymbol{\beta} \in  \mathbb{R}^p$. Define
\begin{equation}
 \label{pitheta}
 \pi(\mathbf{X}, \boldsymbol\theta)=(1-\varepsilon)\phi(\boldsymbol{\beta}^T \mathbf{X})+\varepsilon.
\end{equation}
Assume that for some $\boldsymbol\theta_0 \in  \mathbb{R}^{p+1}$, $\pi(\textbf{X})=\pi(\textbf{X},\boldsymbol\theta_0)$.
We propose to estimate the mean of $Y$ by
\begin{equation}
 \label{ipw_missinga}
\hat\mu_{P}=\mathbb P_n\left\{\frac{AY}{ \pi(\textbf{X},\hat{ \boldsymbol{\theta}}_n)}\right\},
\end{equation}
where $\hat{\boldsymbol{\theta}}_n$ is
the maximum likelihood estimator of  $\boldsymbol{\theta}_0$ under model (\ref{pitheta}).

In order to identify the parameter $\boldsymbol\theta_0$  from the distribution of  $(\textbf{X}, A)$, the following assumptions are required:
\label{supuestosmissing}
\begin{enumerate}
\item[\textbf{A.1}]  The function $\phi$ is injective.
\item[ \textbf{A.2}]  The distribution of the observed covariates is not concentrated at a hyperplane.
\item[\textbf{A.3}]  For all $\boldsymbol{\beta} \in \mathbb{R}^{p}$ such that $ A\,|\,\mathbf X \sim A\,|\, \boldsymbol{\beta}^T\mathbf X$, the support of $\boldsymbol{\beta}^T\mathbf X$ is unbounded from below:
\begin{equation*}
\inf \left\{z \in \mathbb R: P(\vert\boldsymbol{\beta}^T\mathbf X -z\vert <r)>0, \hbox{for all $r>0$}\right\}=-\infty.
\end{equation*}
\end{enumerate}

While the first two assumptions are the classic requirements to obtain the identifiability of the parameter
given in a GLM model as (\ref{para}), the last assumption guarantees the identifiability of the parameters involved in the model introduced in this work.\\
Assume that  \textbf{A.1}, \textbf{A.2} and \textbf{A.3} hold. Under
regularity conditions, it can be proved that
$\hat{\boldsymbol\theta}_n$  is a consistent estimator of $\boldsymbol \theta_0$, 
and therefore $\hat \mu_P\,$ is a
consistent estimator of $\mu$.

\section{Average Treatment effect}\label{sec:ate}
Causal Inference is a second field where inverse probability weighted procedures play a crucial role.
Consider, for instance,  a dichotomous treatment
variable $T$, where $T=1$ represents an active treatment  and $T=0$
means that a  control is assigned. The potential outcomes framework,  introduced by Neyman \cite{neyman23} and Rubin \cite{rubin74}, is used to  quantify the effect of the treatment on some response of interest, whenever this difference is different from zero.
To do so, two potential outcomes (or counterfactual variables)
$Y^{(0)}$ and $Y^{(1)}$  are defined to  represent the   outcome variable of interest that would be seen if an individual   were to receive the treatment and the control, respectively.
 We are interested in estimating the average treatment effect (ATE),
defined as the difference between the mean values of the potential outcomes:
$\tau= E(Y^{(1)})-E(Y^{(0)})$.
$E(Y^{(1)})$ (respectively $E(Y^{(0)})$) represents the hypothetical mean of the response for the population of individuals where all of them  are assumed to  receive treatment (respectively control). So, the difference between these means may be considered a resultant of the treatment, meaning that it  has a  \textit{causal effect} on the response of interest, whenever this difference is different from zero.

The potential outcomes  $Y^{(1)}$ and  $Y^{(0)}$  constitute an artificial
contraption that allows us to conceptualize
what we mean by \textit{causality}.
Only one of these variables is observed in each individual and  it is  related to the observed response through the consistency assumption which
establishes (see Reference \cite{colefrangakis09})  that, if an individual follows the treatment ($T=1$), then the potential outcome $Y^{(1)}$ is precisely his observed outcome. It also establishes the same thing for the control case.
Therefore, under the consistency assumption,
the observed outcome $Y$ is related to the counterfactual variables through the  identity $Y=TY^{(1)}+(1-T)Y^{(0)}$.
Thus, estimating the average treatment  effect is a missing data problem, since one of the counterfactuals ($Y^{(1)}$ or  $Y^{(0)}$)  is missing for each  individual.
To identify $\tau$  we assume  that a vector $\boldsymbol X$ with \textit{all  possible counfounders}  is observed at each subject. This means  that   potential outcomes  $Y^{(1)}$ and $Y^{(0)}$ are conditionally independent of the
treatment exposure $T$ given $\boldsymbol X$:
\begin{equation}
\label{NUC}
 (Y^{(0)}, Y^{(1)}) \coprod T \vert \boldsymbol X.
\end{equation}
This assumption is known in the literature as strongly ignorable treatment
assignment or no unmeasured confounders (see Reference \cite{rosenbaumrubin83}).
Assuming (\ref{NUC}),   we have that
\begin{equation}
\label{identification}
E\left(Y^{(1)}\right)=E\left\{\frac{TY}{\pi(\boldsymbol X)}\right\} \;,\quad E\left(Y^{(0)}\right)=E\left\{\frac{(1-T)Y}{1-\pi(\boldsymbol X)}\right\},
\end{equation}
where now  the propensity score $\pi(\boldsymbol X)$ is defined by  $\pi(\boldsymbol X)=P(T=1\mid \boldsymbol X)$.
These representations of $E\left(Y^{(1)}\right)$ and $E\left(Y^{(0)}\right)$ immediately suggest the estimator
for the ATE  proposed by Rosenbaum \cite{rosenbaum98}, given by
\begin{equation}
 \label{est.gen}
 \hat \tau(\hat \pi)= \mathbb P_n \left\{\frac{TY}{\hat \pi(\textbf{X})}\right\} -\mathbb P_n \left\{\frac{ (1-T)Y}{1-\hat \pi(\textbf{X})}\right\},
\end{equation}
where  $\hat \pi(\textbf{X})$ is a consistent estimator of $\pi(\textbf{X})$.
Lunceford and Davidian \cite{lunceforddavidian04}, following the general framework of \cite{robinsetal94}
and  the  theory of M-estimation (see Reference \cite{stefanskiboos02}),  presented large-sample  theoretical properties of this estimator when $\pi$ follows a linear logistic model.
 Yao, Sun and Wang \cite{yaoetal10} postulate a generalized linear model for the propensity score, as in (\ref{para}),  and  estimate $\pi(\boldsymbol X)$  with
 $ \phi\left(\hat{\boldsymbol{\beta}}_n^T \boldsymbol X\right)$, where $\hat {\boldsymbol{\beta}}_n$ denotes the MLE under model (\ref{para}).
Non parametric versions of $(\ref{est.gen})$ are presented and analysed by Hirano, Imbens  and Ridder  \cite{hiranoetal03}, who propose splines in order to estimate $\pi(\textbf{X})$. Their estimator achieves the semiparametric efficiency bound
established by Hahn \cite{hahn98}.
Both $\hat \pi(\boldsymbol X)^{-1}$ and $\{1-\hat \pi(\boldsymbol X)\}^{-1}$ are now involved in the estimator presented in (\ref{est.gen}). Thus,
 in this scenario, the strict positivity assumption is restated in terms of a lower and  an upper bound for the propensity score, which is now assumed to be bounded away from zero and one,  in the sense that $P\left\{\varepsilon \leq \pi(\boldsymbol X)\leq 1-\delta\right\}=1$, for some $\varepsilon$ and $\delta$ greater than zero. This assumption, also  known as (existence of)  \textit{overlap} in the covariate distributions, is usually assumed to derive the asymptotic properties of $\hat \tau(\hat \pi)$, for the different estimators $\hat{\pi}$ of $\pi$ considered in the literature.  See References \cite{robinsetal95, hiranoetal03,lunceforddavidian04,crumpetal09}. Besides, the lack of overlap  leads to an erratic behavior of $\hat \tau(\hat\pi)$, making the  precise estimation  of  $\tau$ difficult.
 To deal with this issue,  some authors propose trimmed   versions   of  (\ref{est.gen}) (see References \cite{colehernan08,crumpetal09,leeetal11,hill13}). 
As already mentioned in Section \ref{sec:missing}, generalized linear models for the propensity score are typically incompatible with the strict positivity assumption.
Thus, using the same ideas developed in the missing data setting,
for $\boldsymbol\theta=(\varepsilon, \delta, \boldsymbol{\beta}^T)^T \in \boldsymbol\theta \subseteq \mathbb{R}^{p+2}$,  where $\varepsilon, \delta \in (0,1)$ with $\varepsilon +\delta <1$ and $\boldsymbol{\beta} \in \mathbb{R}^p,$  we propose the following model
\begin{equation}
 \label{propate}
 \pi(\textbf{X}, \boldsymbol\theta)= (1-\delta-\varepsilon)\phi(\boldsymbol{\beta}^T \textbf{X})+\varepsilon
\end{equation}
and assume that
$\pi(\textbf{X})=\pi(\textbf{X},\boldsymbol\theta_0)$,  for some $\boldsymbol\theta_0\in \boldsymbol\theta.$  Model (\ref{propate})  intends to preserve as much as possible from the original family postulated for the propensity ($\phi(\boldsymbol{\beta}^T \textbf{X})$); the proposed modification contemplates the strict positivity condition by the inclusion of $\varepsilon$ and $\delta$.   Model (\ref{propate})  will be called
strictly positive propensity score (SPPS) for treatment assignment,  or simply SPPS model if the context is clear. So, the estimator for the ATE that we consider under model (\ref{propate}) for the propensity score  is given by
\begin{equation}
\label{nuestroate}
 \hat \tau_{P}= \mathbb P_n \left\{\frac{TY}{ \pi(\textbf{X}, \hat{\boldsymbol\theta}_n)}\right\}-\mathbb P_n \left\{ \frac{(1-T)Y}{1- \pi(\textbf{X},  \hat{\boldsymbol\theta}_n)}\right\},
\end{equation}
where $ \hat{\boldsymbol\theta}_n$ is the maximum likelihood estimator of  $\boldsymbol\theta_0$ under model (\ref{propate}).\\

Lunceford and Davidian \cite{lunceforddavidian04} propose a modified IPW estimator for $\tau$, in which the weights in $\hat\tau(\hat\pi)$ are redefined in such a way that they not do not take on extremely large values.
The authors  show that this modification gives rise to a more stable procedure. We apply the same modification to the weights computed assuming the SPPS model and obtain the following estimator for the ATE, which combines the proposal of Lunceford and Davidian \cite{lunceforddavidian04} and the one proposed in this work.\\
\begin{eqnarray}
\nonumber
 \hat \tau_{LDP}=\left ( \mathbb P_n \left[\dfrac{T}{{\pi}(\textbf{X},\hat{\boldsymbol\theta}_n)}\left\{ 1- \dfrac{C_1}{{\pi}(\textbf{X}, \hat{\boldsymbol\theta}_n)}\right\} \right]\right)^{-1}   \mathbb P_n \left[\dfrac{TY}{ \pi(\textbf{X}, \hat{\boldsymbol\theta}_n)}\left\{ 1- \dfrac{C_1}{{\pi}(\textbf{X},\hat{\boldsymbol\theta}_n)}\right\} \right]
\end{eqnarray}
\begin{eqnarray}
\label{estilucen}
    - \left ( \mathbb P_n \left[\dfrac{1-T}{1- {\pi}(\textbf{X},\hat{\boldsymbol\theta}_n)}\left\{ 1- \dfrac{C_0}{1-{\pi}(\textbf{X},\hat{\boldsymbol\theta}_n)}\right\}  \right]\right)^{-1}   \mathbb P_n \left[\dfrac{(1-T) Y}{1- \pi(\textbf{X},\hat{\boldsymbol\theta}_n)}\left\{ 1- \dfrac{C_0}{1-{\pi}(\textbf{X},\hat{\boldsymbol\theta}_n)}\right\} \right]
\end{eqnarray} \normalsize
where
$$ C_1= \dfrac{ \mathbb P_n \left\{ \dfrac{T-{\pi}(\textbf{X},\hat{\boldsymbol\theta}_n)}{{\pi}(\textbf{X},\hat{ \boldsymbol\theta}_n)} \right\}
}{ \mathbb P_n \left[ \left \{\dfrac{T-{\pi}(\textbf{X},\hat{\boldsymbol\theta}_n)}{{\pi}(\textbf{X},\hat{\boldsymbol\theta}_n)}\right\}^2 \right]},\,\,\,\; C_0= - \dfrac{ \mathbb P_n \left\{ \dfrac{T-{\pi}(\textbf{X},\hat{\boldsymbol\theta}_n)}{1-{\pi}(\textbf{X},\hat{ \boldsymbol\theta}_n)} \right\}}{ \mathbb P_n \left[ \left \{\dfrac{T-{\pi}(\textbf{X},\hat{\boldsymbol\theta}_n)}{1-{\pi}(\textbf{X},\hat{\boldsymbol\theta}_n)}\right\}^2 \right]},$$
\normalsize
and $ \hat{\boldsymbol\theta}_n$ is the maximum likelihood estimator of  $\boldsymbol\theta_0$ under model (\ref{propate}).
 
In this setting, in order to identify the parameter $\boldsymbol\theta_0$ and in addition to \textbf A.1, \textbf A.2 and \textbf A.3 we need the following assumption.
\begin{enumerate}
\item[\textbf A.4]  For all $\boldsymbol{\beta} \in \mathbb{R}^{p}$ such that $ A\,|\,X \sim A\,|\, \boldsymbol{\beta}^T \mathbf X,$ the support of $\boldsymbol{\beta}^T\mathbf X$ is unbounded from above:
\begin{equation*}
\sup \left\{z \in \mathbb R: P(\vert\boldsymbol{\beta}^T\mathbf X -z\vert <r)>0, \hbox{for all $r>0$}\right\}=+\infty.
\end{equation*}
\end{enumerate}
Under regularity conditions and assumptions \textbf{A.1}, \textbf{A.2},  \textbf{A.3} and \textbf{A.4} , it can be proved that $\boldsymbol\theta _0$ is identifiable and consistently estimated by   $\hat{\boldsymbol\theta}_n$, and so, $\hat \tau_P$ is a
consistent estimator of $\tau$.\\
\section{Computational Method for Fitting the SPPS Model}
In this section we describe the computational method used for the maximum likelihood estimation of the $p+1$ parameters in the SPPS model for the missing mechanism and the $p+2$ parameters in the SPPS model for treatment assignment.
We describe in detail the method we propose for estimation in the treatment assignment case, giving in between brackets the necessary modifications for the missing data setting.
Given $(\mathbf X_1,T_1)\dots (\mathbf X_n,T_n)$ a random sample from model (\ref{propate}) described in  Section \ref{sec:ate}, we estimate $\varepsilon,\delta$ and $\boldsymbol{\beta}$ iteratively with the following procedure: consider that $T_i|\mathbf X_i$ follows a Bernoulli distribution with $P(T_i=1 |\mathbf X_i)=\pi(\textbf{X}_i, \boldsymbol\theta_0)$  where $\pi$ is given by equation (\ref{propate}). Estimate iteratively $\boldsymbol\beta$ assuming $\varepsilon$ and $\delta$ are known, $\varepsilon$ assuming $\boldsymbol\beta$ and $\delta$ are known, and $\delta$ assuming $\boldsymbol\beta$ and $\varepsilon$ are known.\\
\begin{enumerate}
\item[Step 1] Substitute $0$ for $\varepsilon$ and $\delta$ in (\ref{propate}) and compute an initial estimate of $\boldsymbol{\beta}$ by maximum likelihood. Call this estimate $\hat{\boldsymbol{\beta}}_0$.
\item[Step 2] Compute the fitted values of the model considered in Step 1, $\hat{\pi_i}= \phi\left( \hat{\boldsymbol{\beta}}_0^T \mathbf X_i\right)$.
\item[Step 3] Compute initial estimates of $\varepsilon$ and $\delta$ by $\hat\varepsilon_1=min\left(\hat\pi_i\right)$ and $\hat\delta_1=1-max\left(\hat\pi_i\right)$. (In the missing data case set  $\hat\delta_1=0$).
\item[Step 4] Substitute $\hat\varepsilon_1$ for $\varepsilon$ and $\hat\delta_1$ for $\delta$ in (\ref{propate}). Then recompute an estimate of $\boldsymbol{\beta}$ by maximum likelihood. Call this estimate $\hat{\boldsymbol{\beta}_1}$.
\item[Step 5] Substitute $\hat{\boldsymbol\beta}_1$ for $\boldsymbol\beta$ and $\hat\delta_1$ for $\delta$ in (\ref{propate}) and estimate $\varepsilon$ by maximum likelihood. This estimate is the new $\hat{\varepsilon_1}$.
\item[Step 6] Substitute $\hat{\boldsymbol\beta}_1$ for $\boldsymbol\beta$ and $\hat\varepsilon_1$ for $\varepsilon$ in (\ref{propate}) and estimate $\delta$ by maximum likelihood. This estimate is the new $\hat{\delta_1}$. (Skip this step in the missing data case).
\item[Step 7] Repeat steps 4 to 6 until convergence (4 and 5 in the missing data case). Return $\hat{\boldsymbol\pi}=(\hat\pi_1,\dots,\hat\pi_n)$.
\end{enumerate}
This iterative method converges rapidly in the case of the missing data setting. In the causality setting it also converges rapidly if the real values of $\varepsilon$ and $\delta$ are such that $\varepsilon+\delta $ is not too large. If this is not the case, the modified link function in the GLM in Step 4 can be very flat, growing very slowly from $\hat\varepsilon_1$ to $1-\hat\delta_1$, which are very similar. When this happens the GLM method in R usually does not converge, except for very large samples. To solve this problem, in the causality setting, we change Step 3  by the following

\begin{enumerate}
\item[Step 3 '] Compute initial estimators of $\varepsilon$ and $\delta$ by $\hat\varepsilon_1=min\left(\hat\pi_i\right)$ and $\hat\delta_1=1-max\left(\hat\pi_i\right)$. If $\hat\varepsilon_1+\hat\delta_1>0.6$, return $\hat{\boldsymbol\pi}=(\hat\pi_1,\dots,\hat\pi_n)$, otherwise proceed to step 4.
\end{enumerate}
We remark that Step 3' implies that, if  after fitting a regular GLM we have that the range of the fitted values is not very large, namely smaller than $0.4$, then we keep the regular GLM; there is no need for the SPPS model. \\
The code that defines the necessary functions for computing the ATE and IPW estimators is available under request.

\section{Monte Carlo Study}
In this section we report the results of a Monte Carlo study we performed in order to assess the advantages of considering the proposed model over the traditional approach.
In practice, several covariates will be available for modeling the propensity score. In order to investigate performance in a realistic setting, Lunceford and Dadivian \cite{lunceforddavidian04}, carried out simulations involving continuous and discrete variables, some of them associated with both treatment exposure and potential response and others associated with potential responses but not treatment exposure. We generate variables as they did in one of their proposed scenarios. We consider covariates $\textbf{X}=(X_1,X_2,X_3,V_1,V_2,V_3),\,$ a binary variable $T\,$ and an outcome $Y\,$ such that the variable $\,T\,$ follows a Bernoulli distribution with
\begin{equation*}
\pi(\textbf{X})=\pi(\textbf{X},\boldsymbol\theta_0)= \varepsilon_0 + \left(1 - \delta_0 - \varepsilon_0  \right) \left\{1 + exp(\boldsymbol{\beta}_0^T \textbf{X} ) \right\},^{-1}
 \end{equation*}
where  $\boldsymbol\theta_0=(\varepsilon_0,\delta_0,\boldsymbol{\beta_0})$ and $\boldsymbol{\beta_0}=(0,0.6,-0.6,0.6,0,0,0)$.\\
  We remark that, when $\varepsilon_0=\delta_0=0$ this SPPS model reduces to the usual linear logistic model.
Different settings of $\varepsilon_0$ and $\delta_0$ will be chosen. The outcome is generated by
  $$ Y=\nu_0 + \nu_1 X_1+\nu_2 X_2+ \nu_3 X_3 + \nu_4 T +  \xi_1 V_1 + \xi_2 V_2 + \xi_3 V_3 + Z,$$
  where $Z$ is a normal standard variable independent of $\mathbf X$ and $T$, $\left( \nu_0, \nu_1,\nu_2,\nu_3,\nu_4 \right)= \left( 0, -1, 1, -1, 2 \right)^T$ and $\xi=(-1, 1, 1)
.$
  The joint distribution of $\mathbf X$ is specified by taking $X_3$ with Bernoulli(0.2) distribution, and then generating $V_3$
as Bernoulli with $P(V_3=1 \vert X_3)= 0.75X_3 + 0.25(1-X_3).$ Conditional on $X_3,$ the vector $\left( X_1,V_1,X_2,V_2\right)$ is generated as multivariate normal $\mathcal N(\rho_{X_3},\Sigma_{X_{3}}),$ where $ \rho_0=\left(1,1,-1,-1 \right),\,\rho_1=\left(-1,-1,1,1 \right),$ and
$$ \Sigma_{0}=\Sigma_1
= A = \left(\begin{array}{cccc}
                          1&0.5& -0.5 & -0.5 \\
                          0.5&1& -0.5 & -0.5 \\
                          -0.5 & -0.5 & 1&0.5 \\
                          -0.5 & -0.5 & 0.5&1
            \end{array}
     \right)$$

We generate $Nrep=1000$ samples of size $n=1000$ following the described model, under which the real value of the average treatment effect is $\tau=2$. For each sample we compute four estimators, namely $\hat\tau_O$, $\hat\tau_P$, $\hat\tau_{LD}$ and $\hat\tau_{PLD}$. $\hat\tau_P$ and $\hat\tau_{PLD}$ are defined in (\ref{nuestroate}  ) and (\ref{estilucen}) respectively; both of them use the SPPS model (\ref{propate}) introduced in this work for the propensity score.  On the other hand, $\hat\tau_O$ is the  original IPW estimator, defined by modeling the propensity score with the linear logistic model. That is to say, $\hat\tau_O=\hat \tau( \hat \pi)$ given in (\ref{est.gen})  assuming that  $\pi(\textbf{X})= \phi(\boldsymbol{\beta}_1^T \textbf{X})$, with $\phi(u) = 1/(1 +exp(-u)) $ for some $\boldsymbol\beta_1 \in  \mathbb{R}^7.$
Finally, $\hat \tau_{LD}$ denotes the estimator proposed in Lunceford and Davidian (2004). This estimator is defined as in equation (\ref{est.gen}) but using a linear logistic regression model for the propensity score.
For each estimator $\hat\tau$, we compute an empirical mean square error with the following formula
\begin{equation*}
{MSE}(\hat\tau) =\frac{1}{Nrep} \sum_{i=1}^{Nrep} \left(\hat\tau-\tau\right)^2.
\end{equation*}

The results of the Monte Carlo study are reported in Table \ref{mse}. In these simulations we can see that, when the true model is the linear logistic model, both our proposed estimators give a smaller MSE than the corresponding classical ones.
On the other hand, if the samples are generated following the SPPS model with $\epsilon_0$ and $\delta_0$ both greater than zero, then $\hat\tau_{PLD}$ remains very small, as compared to the other proposals, followed sometimes by $\hat\tau_{P}$ and sometimes by $\hat\tau_{LD}$. We remark that in almost all the situations considered our proposed estimators give better results than the corresponding classical ones. In the remaining situations, namely when samples are generated with large values of $\epsilon_0$ and $\delta_0$, the mean square errors are similar and all the estimators give good results.  When $\epsilon_0+\delta_0$ is greater than $0.6$, all four estimators are equal or almost equal. \\

\section{Example: Children's FEV Data}
This data set contains measurements of the forced expiratory volume  (FEV) of 654 children and teenagers aged 3 to 19 years, together with their height, age, sex and and a binary variable indicating whether or not they smoke. It is basically the data set considered in \cite{rosner99}, which has been included in the R package covreg \cite{niuhoff14} 
with slight modifications.
In order to estimate the average smoking effect in the FEV in this population, we consider only children aged 9 or older since there are not any smokers among the younger children in this data set. 

We wish to emphasize that this example is included to show the use of the proposed model in a causal inference context. The confiability of  results  depend on the no unmeasured counfounders assumption, and  the discussion on its validity given the observed covariates is beyond the scope of this work. This example is therefore considered only as an illustration of the application of the proposed methods. 

All four estimates yield a negative average treatment effect but differ in the size of this effect. 
As a means to determine the significance of these differences, we computed $95\%$ normal confidence intervals were we estimated the standard error based on 1000 bootstrap samples.
 To obtain the bootstrap samples, we first break the data set in two, smokers (65 subjects) and non-smokers (374 subjects). Then we resample separately 65 observations from the smokers group and 374 observations from the non smokers group. Let $\hat\tau_i$ be the estimator of the ATE based on the $ith$ sample and $\overline{\hat{\tau}}$ the mean of $\hat\tau_1,\dots \hat\tau_{Nboot}$. The bootstrap estimator of the standard error of each estimator $\hat\tau$ is computed by
\begin{equation}
\hat{SE}=\sqrt{\frac{1}{Nboot-1}\sum_ {i=1}^{Nboot} \left(\hat\tau_i-\overline{\hat{\tau}}\right)^2}
\end{equation}
and the $95\%$ bootstrap confidence interval based on $\hat\tau$ is computed as \begin{equation}
\left[ \hat\tau-1.96\ \ \hat{SE} ,\ \  \hat\tau+1.96\ \ \hat{SE} \right]
\end{equation}

As reported in Table \ref{tablesmoke}, the obtained intervals show that the  ATE is not significant. The conclusion is the same for all the estimators considered.

 In Table \ref{tablesmoke} we report the estimates, together with their corresponding standard errors and confidence interval for the ATE.
 We remark that the standard error of our proposed estimator $\hat{\tau}_P$ is slightly smaller than the standard errors of the classical ones.

 \bigskip

\textbf{Acknowledgement} 
This research was partly supported by grants 20020150200110BA and 20020130100279BA from Universidad de Buenos Aires.

 \newpage

\begin{center}
 \begin{table}{ht}
 \centering
\begin{tabular}{cc|c|c|c|c|c|c|c}
\hline
$\delta | \epsilon$&&0&0.05&0.1&0.2&0.3&0.4&0.5\\ \hline
0 &$\begin{array}{l} \hat\tau_O\\ \hat\tau_P \\ \hat\tau_{LD} \\ \hat\tau_{PLD} \end{array}$
&\begin{tabular}{r} 0.204 \\ 0.109 \\ 0.058 \\ 0.055 \end{tabular}
&\begin{tabular}{r} 0.138 \\ 0.07 \\ 0.045 \\ 0.037 \end{tabular}
&\begin{tabular}{r} 0.094 \\ 0.065 \\ 0.041 \\ 0.035 \end{tabular}
&\begin{tabular}{r} 0.075 \\ 0.069 \\ 0.046 \\ 0.038 \end{tabular}
&\begin{tabular}{r} 0.109 \\ 0.079 \\ 0.066 \\ 0.044 \end{tabular}
&\begin{tabular}{r} 0.156 \\ 0.099 \\ 0.098 \\ 0.054 \end{tabular}
&\begin{tabular}{r} 0.198 \\ 0.125 \\ 0.133 \\ 0.073 \end{tabular}
\\ \hline
0.05 &$\begin{array}{l} \hat\tau_O\\ \hat\tau_P \\ \hat\tau_{LD} \\ \hat\tau_{PLD} \end{array}$
&\begin{tabular}{r} 0.221 \\ 0.066 \\ 0.048 \\ 0.04 \end{tabular}
&\begin{tabular}{r} 0.176 \\ 0.036 \\ 0.045 \\ 0.026 \end{tabular}
&\begin{tabular}{r} 0.111 \\ 0.031 \\ 0.041 \\ 0.023 \end{tabular}
&\begin{tabular}{r} 0.052 \\ 0.028 \\ 0.031 \\ 0.022 \end{tabular}
&\begin{tabular}{r} 0.041 \\ 0.028 \\ 0.029 \\ 0.022 \end{tabular}
&\begin{tabular}{r} 0.047 \\ 0.03 \\ 0.034 \\ 0.024 \end{tabular}
&\begin{tabular}{r} 0.053 \\ 0.035 \\ 0.041 \\ 0.028 \end{tabular}
\\ \hline
0.1 &$\begin{array}{l} \hat\tau_O\\ \hat\tau_P \\ \hat\tau_{LD} \\ \hat\tau_{PLD} \end{array}$
&\begin{tabular}{r} 0.171 \\ 0.06 \\ 0.047 \\ 0.039 \end{tabular}
&\begin{tabular}{r} 0.149 \\ 0.03 \\ 0.044 \\ 0.023 \end{tabular}
&\begin{tabular}{r} 0.104 \\ 0.026 \\ 0.042 \\ 0.021 \end{tabular}
&\begin{tabular}{r} 0.047 \\ 0.023 \\ 0.029 \\ 0.02 \end{tabular}
&\begin{tabular}{r} 0.029 \\ 0.022 \\ 0.022 \\ 0.019 \end{tabular}
&\begin{tabular}{r} 0.026 \\ 0.022 \\ 0.022 \\ 0.02 \end{tabular}
&\begin{tabular}{r} 0.026 \\ 0.024 \\ 0.023 \\ 0.021 \end{tabular}
\\ \hline
0.2 &$\begin{array}{l} \hat\tau_O\\ \hat\tau_P \\ \hat\tau_{LD} \\ \hat\tau_{PLD} \end{array}$
&\begin{tabular}{r} 0.088 \\ 0.061 \\ 0.054 \\ 0.04 \end{tabular}
&\begin{tabular}{r} 0.075 \\ 0.029 \\ 0.036 \\ 0.023 \end{tabular}
&\begin{tabular}{r} 0.058 \\ 0.024 \\ 0.032 \\ 0.02 \end{tabular}
&\begin{tabular}{r} 0.029 \\ 0.02 \\ 0.022 \\ 0.018 \end{tabular}
&\begin{tabular}{r} 0.019 \\ 0.02 \\ 0.017 \\ 0.019 \end{tabular}
&\begin{tabular}{r} 0.015 \\ 0.018 \\ 0.014 \\ 0.017 \end{tabular}
&\begin{tabular}{r} 0.015 \\ 0.016 \\ 0.015 \\ 0.015 \end{tabular}
\\ \hline
0.3 &$\begin{array}{l} \hat\tau_O\\ \hat\tau_P \\ \hat\tau_{LD} \\ \hat\tau_{PLD} \end{array}$
&\begin{tabular}{r} 0.097 \\ 0.071 \\ 0.078 \\ 0.047 \end{tabular}
&\begin{tabular}{r} 0.045 \\ 0.031 \\ 0.034 \\ 0.026 \end{tabular}
&\begin{tabular}{r} 0.032 \\ 0.025 \\ 0.025 \\ 0.022 \end{tabular}
&\begin{tabular}{r} 0.018 \\ 0.02 \\ 0.016 \\ 0.018 \end{tabular}
&\begin{tabular}{r} 0.013 \\ 0.017 \\ 0.013 \\ 0.016 \end{tabular}
&\begin{tabular}{r} 0.012 \\ 0.013 \\ 0.012 \\ 0.013 \end{tabular}
&\begin{tabular}{r} 0.012 \\ 0.012 \\ 0.012 \\ 0.012 \end{tabular}
\\ \hline
0.4 &$\begin{array}{l} \hat\tau_O\\ \hat\tau_P \\ \hat\tau_{LD} \\ \hat\tau_{PLD} \end{array}$
&\begin{tabular}{r} 0.166 \\ 0.085 \\ 0.118 \\ 0.057 \end{tabular}
&\begin{tabular}{r} 0.05 \\ 0.037 \\ 0.04 \\ 0.032 \end{tabular}
&\begin{tabular}{r} 0.028 \\ 0.028 \\ 0.025 \\ 0.025 \end{tabular}
&\begin{tabular}{r} 0.017 \\ 0.02 \\ 0.016 \\ 0.019 \end{tabular}
&\begin{tabular}{r} 0.012 \\ 0.013 \\ 0.012 \\ 0.013 \end{tabular}
&\begin{tabular}{r} 0.011 \\ 0.011 \\ 0.011 \\ 0.011 \end{tabular}
&\begin{tabular}{r} 0.012 \\ 0.012 \\ 0.012 \\ 0.012 \end{tabular}
\\ \hline
0.5 &$\begin{array}{l} \hat\tau_O\\ \hat\tau_P \\ \hat\tau_{LD} \\ \hat\tau_{PLD} \end{array}$
&\begin{tabular}{r} 0.264 \\ 0.101 \\ 0.178 \\ 0.067 \end{tabular}
&\begin{tabular}{r} 0.061 \\ 0.044 \\ 0.049 \\ 0.038 \end{tabular}
&\begin{tabular}{r} 0.03 \\ 0.03 \\ 0.028 \\ 0.028 \end{tabular}
&\begin{tabular}{r} 0.017 \\ 0.017 \\ 0.017 \\ 0.017 \end{tabular}
&\begin{tabular}{r} 0.013 \\ 0.013 \\ 0.013 \\ 0.013 \end{tabular}
&\begin{tabular}{r} 0.012 \\ 0.012 \\ 0.012 \\ 0.012 \end{tabular}
&\begin{tabular}{r} 0.011 \\ 0.011 \\ 0.011 \\ 0.011 \end{tabular}
\\ \hline
\end{tabular}
\normalsize
\caption{Empirical mean square errors of different estimators of ATE}
\label{mse}
\end{table}
\end{center}

\begin{table}[ht]
\centering
\begin{tabular}{lrrrr}
  \hline
 & Estimator & Standard Error & Confidence Interval \\
  \hline
$\hat \tau_O$ & -0.1538 & 0.2188 & (-0.5826 , 0.2750) \\
$\hat \tau_P$ & -0.2663 & 0.1952 & (-0.6727 , 0.1402) \\
$\hat \tau_{LD}$ & -0.1911 & 0.2074 & (-0.5737 , 0.1914)\\
$\hat \tau_{PLD}$ & -0.2637 & 0.2086 & (-0.6726 , 0.1452) \\
   \hline
\end{tabular}
\caption{Estimates for the ATE of smoking in children, their standard errors and confidence intervals.}
\label{tablesmoke}
\end{table}


\begin{thebibliography}{99}

\bibitem{robinsetal94} Robins JM, Rotnitzky A, Zhao LP. Estimation of regression coefficients when some regressors are not always observed. Journal of the American statistical Association 1994; 89(427): 846-866.

\bibitem{robinsetal95} Robins JM, Rotnitzky A, Zhao LP. Analysis of semiparametric regression models for repeated outcomes in the presence of missing data. Journal of the American Statistical Association 1995; 90(429): 106-121.

\bibitem{littlean04} Little R, An H. Robust likelihood-based analysis of multivariate data with missing values. Statistica Sinica 2004; 949-968.

\bibitem{kangshafer07} Kang JD, Schafer JL. Demystifying double robustness: A comparison of alternative strategies for estimating a population mean from incomplete data. Statistical science 2007; 523-539.

\bibitem{rosenbaum98} Rosenbaum PR. Propensity score. Encyclopedia of Biostatistics, Armitage P, Colton T (eds), vol. 5. Wiley: New York, 1998; 3551–3555. 


\bibitem{hiranoetal03} Hirano K, Imbens GW, Ridder G. Efficient estimation of average treatment effects using the estimated propensity score. Econometrica 2003; 71(4): 1161-1189.


\bibitem{lunceforddavidian04} Lunceford JK, Davidian M. Stratification and weighting via the propensity score in estimation of causal treatment effects: a comparative study. Statistics in medicine 2004; 23 (19): 2937-2960.

\bibitem{crumpetal09} Crump RK, Hotz VJ, Imbens GW, Mitnik OA. Dealing with limited overlap in estimation of average treatment effects. Biometrika 2009; 96(1): 187-199.


\bibitem{yaoetal10}  Yao L, Sun Z, Wang Q. Estimation of average treatment effects based on parametric propensity score model. Journal of Statistical Planning and Inference 2010; 140 (3): 806-816.

\bibitem{hill13} Hill JB, Robust Estimation for Average Treatment Effects (February 26, 2013). Available at SSRN: https://ssrn.com/abstract=2260573 or http://dx.doi.org/10.2139/ssrn.2260573

\bibitem{littlerubin87} Little RJ, Rubin DB. Statistical Analysis with Missing Data. John Wiley and Sons 1987.

\bibitem{colehernan08} Cole SR, Hernan MA. Constructing inverse probability weights for marginal structural models. American journal of epidemiology 2008; 168(6): 656-664.

 \bibitem{leeetal11} Lee BK, Lessler J, Stuart E A. Weight trimming and propensity score weighting. PloS one 2011; 6(3): e18174.

\bibitem{tageretal79} Tager IB, Weiss A, Rosner B, Speizer, FE. Effect of parental cigarette smoking on the pulmonary function of children. American Journal of Epidemiology 1979; 110 (1): 15-26.

\bibitem{tageretal83} Tager IB, Weiss ST, Muñoz A, Rosner B, Speizer FE. Longitudinal study of the effects of maternal smoking on pulmonary function in children. New England Journal of Medicine 1983; 309 (12): 699-703.

\bibitem{rosner99} Rosner B. Fundamentals of Biostatistics, 5th ed., Pacific Grove, CA: Duxbury 1999.

\bibitem{kahn05} Kahn M. An exhalent problem for teaching statistics. The Journal of Statistical Education 2005; 13 (2).


\bibitem{rubin76} Rubin DB. Inference and missing data. Biometrika 1976; 63 (3): 581–592.


\bibitem{rosenbaumrubin83} Rosenbaum PR, Rubin DB. The central role of the propensity score in observational studies for causal effects. Biometrika  1983; 41-55.

\bibitem{horvitzthompson52} Horvitz DG, Thompson DJ. A generalization of sampling without replacement from a finite universe. Journal of the American statistical Association 1952; 47(260): 663-685.

\bibitem{robinsrotnitzky92} Robins JM, Rotnitzky A. Recovery of information and adjustment for dependent censoring using surrogate markers. Aids Epidemiology. Springer 1992; 297-331


\bibitem{neyman23} Neyman J. Sur les applications de la théorie des probabilités aux experiences agricoles: Essai des principes. Roczniki Nauk Rolniczych 1923; 10: 1-51.


\bibitem{rubin74} Rubin DB. Estimating causal effects of treatments in randomized and nonrandomized studies. Journal of educational Psychology 1974; 66(5): 688.


\bibitem{colefrangakis09} Cole SR, Frangakis CE. The consistency statement in causal inference: a definition or an assumption?. Epidemiology 2009; 20(1): 3-5.


\bibitem{stefanskiboos02} Stefanski LA, Boos DD. The calculus of M-estimation. The American Statistician 2002; 56(1): 29-38.


\bibitem{hahn98} Hahn J. On the role of the propensity score in efficient semiparametric estimation of average treatment effects. Econometrica 1998; 315-331.


\bibitem{niuhoff14} Niu X, Hoff P. covreg: A simultaneous regression
  model for the mean and covariance. R package version 2014.\end{thebibliography}
\end{document}